\documentstyle{article}

\begin{document}
\begin{center} {\Large {\bf
The dipole coupling of atoms and light in gravitational
fields }}
\footnote{To appear in Phys. Rev. A.}\\[1cm]
Karl-Peter Marzlin \\[2mm]
Fakult\"at f\"ur Physik
der Universit\"at Konstanz\\
Postfach 5560 M 674\\
D-78434 Konstanz, Germany
\footnote{e-mail: Peter.Marzlin@uni-konstanz.de}
\end{center} $ $\\[3mm]
\begin{minipage}{15cm}
\begin{abstract}
The dipole coupling term between a system of N particles with
total charge zero and the electromagnetic field is derived
in the presence of a weak gravitational field. It is shown that
the form of the coupling remains the same as in flat
space-time if it is written with respect to the proper time
of the observer and to the measurable field components.
Some remarks concerning the connection between the minimal
and the dipole coupling are given.
\end{abstract} \end{minipage}$ $ \\[3mm]
\section{Introduction}
Although modern theories of quantized matter in curved
space make predictions for extreme situations like, e.g., the
very early universe still very few experiments are done
to shed light on the connection between general relativity
and quantum mechanics.
The recent progress in atomic interferometry
may lead to new contributions on this topic.
For example, the Sagnac
phase \cite{riehle91} and the influence of the earth's
acceleration \cite{kachu91} were measured showing the behaviour
of atoms in non inertial frames of reference. These
experiments, which use the interaction between lasers and atoms
to split and recombine the atomic beam, may also be of use to
measure the influence of space-time curvature on atoms
\cite{maau93b}. It is therefore of interest to study the
behaviour of such devices in weak gravitational fields like
that of the earth. In order to do so it is necessary to
generalize the theoretical methods which are used for the
description of atoms and lasers from a flat space-time to a
curved one.

In quantum optics it is often convenient to use the
dipole coupling  $ - \vec{d} \cdot \vec{E}$ instead of the
minimal coupling scheme in the calculations. While the latter
is invariant under Lorentz transformations the dipole coupling
has the advantage that it is directly related to the physical
electric field $\vec{E}$ and not to the gauge dependent vector
potential $A_{\mu}$. The (non relativistic) equivalence of the
two approaches was first demonstrated by M. G\"oppert-Mayer
\cite{goeppert31}. She showed that the classical Lagrangians
of the two theories are related by a canonical transformation.
Power and Zienau \cite{pozie59} have extended her work
by use of a unitary transformation in quantum theory and
derived, in an approximation, a multipolar Hamiltonian where the
dipole coupling is only the first order interaction term of a
multipole series. This transformation was made exact by
Woolley (see, e.g., Ref. \cite{woolley75}).
All these derivations are formulated in a fixed non-covariant
gauge, mostly the Coulomb gauge. The generalization to an
arbitrary gauge fixing was done by Power and Thirunamachandran
\cite{pothi80} by adding a total time derivative to the
classical Lagrangian and by Woolley \cite{woolley80}.

Despite the formal equivalence of both couplings there has
been a long discussion in the literature which one is better
suited to describe the interaction between matter and light.
Lamb \cite{lamb52} stated that the dipole interaction
is preferable because the results of certain calculations fit
better to the experiment. The minimal coupling
scheme on the other hand is covariant and connected with a
gauge symmetry, facts which seem to imply that this coupling
is more fundamental. Ackerhalt and Milonni
\cite{ackerhalt84} have pointed out that this controversy
may arise from the necessity to transform also the states
if one applies a unitary transformation. This leads to the
question for which coupling scheme the textbook wavefunctions
are the right choice. We will argue in this paper that the
textbook wavefunctions seem to belong to the dipole coupling.

The main subject however is to study the modification of the
dipole coupling in weak gravitational fields.
Our strategy is to begin with
a general relativistic Lagrangian for a multiparticle
system in an electromagnetic field, to consider the limit of
small velocities, and to perform the Power-Zienau transformation
along the lines of Ref. \cite{pothi80}. This approach has two
advantages: one can avoid to use the Coulomb gauge which is not
covariant, and one circumvents the inclusion of gauge
constraints like, e.g., the Gupta-Bleuler condition which are
not common in quantum optics. One disadvantage is that the
spin has to be treated separately. To the knowledge of the
author the dipole coupling in gravitational fields
was addressed only by Bord\'e {\em et al.}
\cite{borde83a,borde83} by replacing the space-indices of the
non-covariant expression by tetrad indices. In this paper it
is the aim to derive its structure from an underlying
multi-particle theory.

We use the conventions of Ref. \cite{MTW} for general
relativity, i.e. sgn $g_{\mu \nu} = +2$.
Greek indices run from 0 to 3 and latin ones from
1 to 3. Summation is understood whenever an index appears twice.
Tetrad indices are underlined.
We use natural units ($\hbar = c =1$)
and Heaviside-Lorentz conventions for the
electromagnetic field.
\section{The derivation of the Hamiltonian}
We begin with the covariant action $S $ of a
system of particles and the electromagnetic field,
\begin{equation} S = \int {\cal L} \sqrt{-g} d^4 x \equiv
    \int L dx^0 \end{equation}
where
\begin{equation} L = - \sum_{\alpha=1}^N \left \{ m_{\alpha}
     \left [ -g_{\mu
     \nu}( x_{(\alpha)} )\; \dot{x}^\mu_{(\alpha)}
     \dot{x}^\nu_{(\alpha)} \right ]^{-1/2} + q_{(\alpha)}
     A_{\mu}(x_{(\alpha)})  \dot{x}^\mu_{(\alpha)} \right \}
     - {1\over 4} \int
     d^3y \sqrt{-g(y)} F_{\mu \nu}(y) F^{\mu \nu}(y)\; .
     \end{equation}
$\alpha $ labels the particles which travel on the
trajectories $x^\mu_{(\alpha)}$. The dot denotes the
derivative with respect to the coordinate time $x^0 =
x^0_{(\alpha)} =y^0$. Following Refs. \cite{dewitt66,papini67}
we perform the limit of small velocities, $|\dot{x}^i_{(\alpha)}
| \ll 1 $, and consider only weak gravitational fields,
$g_{\mu \nu} = \eta_{\mu \nu} + h_{\mu \nu}$ with $ |h_{\mu \nu}
| \ll 1$. This leads to the new Lagrangian
\begin{eqnarray} L &=& \sum_{\alpha} \Bigg \{ {m_{(\alpha)}
     \over 2} h_{00} + m_{(\alpha)}\, h_{0i}\, \dot{
     x}^i_{(\alpha)} + {m_{(\alpha)} \over 2} \left [(1+{1\over
     2} h_{00})\delta_{kl} + h_{kl}\right ]
     \dot{x}^k_{(\alpha)}\dot{x}^l_{(\alpha)} \nonumber \\
     & & + q_{(\alpha)}\, A_0 (x_{(\alpha)}) +q_{(\alpha)}\,
     A_i(x_{(\alpha)})\, \dot{x}^i_{(\alpha)} \Bigg \}
     + {1\over 2} \int d^3 y \Bigg \{ \left [ (1+{1\over 2}
     h_{\lambda \lambda}) \delta_{kl} -h_{kl} \right ] \times
     \label{weaklag} \\ & &
     (A_{k,0} - A_{0,k})(A_{l,0}-A_{0,l}) - \left [ (1-{1\over
     2} h_{\lambda \lambda}) \delta_{kl} +h_{kl} \right ]
     B_k B_l - 2(A_{i,0} - A_{0,i}) \varepsilon_{ijk} B_j
     h_{ok}\Bigg \} \; . \nonumber  \end{eqnarray}
Each factor of $h_{\mu \nu}$ in the sum over $\alpha$ has to be
taken at the point $x_{(\alpha)}$. Throughout the paper all
expressions  are calculated to first order in $h_{\mu \nu}$
only. In the derivation of
(\ref{weaklag}) we have subtracted the total rest energy $
M = \sum_{\alpha} m_{(\alpha)} $ and have defined $B_i :=
\varepsilon_{ijk} A_{k,j}$ where $\varepsilon_{ijk}$ is the
total antisymmetric symbol with $\varepsilon_{123} = 1$.
A comma denotes the derivative
with respect to the following coordinate. Here it is
necessary to give a remark concerning the position of the
indices. In general relativity an index which appears twice
has to appear as one upper and one lower index.
Any expression which does not fulfill this
requirement cannot be invariant under coordinate
transformations. Although the starting point of this calculation
was a covariant expression we will use extensively
differential geometric methods in flat three-dimensional
space. It is therefore convenient to switch to a three-space
notation where the indices of any three-vector
$\vec{V} \equiv \{ V_i \}$ are lower case indices except for
all coordinates $x_{(\alpha)} , y , \ldots $ where,
e.g., $\vec{y} \equiv \{ y^i \}$. This can be done without
making errors as long as no index is moved with the
space-time metric, and as long as we do not transform
to another coordinate system. The resulting Hamiltonian
then describes the time evolution of the particles and fields
in this coordinate system.

Eq. (\ref{weaklag}) is the Lagrangian of non relativistic
particles moving in weak gravitational field. If one
defines a Hamiltonian $H = p \dot{x} -L $ one ends up with
the minimal coupling scheme. As will be shown below this
non-covariant way to define H is only justified
in the case of time independent gravitational fields.
In order to derive the dipole approximation of the coupling
we follow Ref. \cite{pothi80} and
add a total time derivative to the Lagrangian:
\begin{equation} L^\prime = L - \frac{d}{dx^0} \int d^3 y
     \vec{P}\cdot \vec{A} \label{switchgl} \end{equation}
with
\begin{equation} \vec{P}(y) = \sum_{\alpha} q_{(\alpha)}
     (\vec{x}_{(\alpha)}
     -\vec{R}_{cm})\, \int_0^1 \delta (\vec{y} -\vec{R}_{cm}
     -\lambda (\vec{x}_{(\alpha)} -\vec{R}_{cm}))\, d \lambda
     \; .\end{equation}
\begin{equation} \vec{R}_{cm} := \frac{1}{M} \sum_{\alpha=1}^N
     m_{\alpha}\vec{x}_{(\alpha)} \end{equation}
is the center of mass of the particles. This is the point where
covariance is explicitly broken by adding a non-covariant
term to $L$. This step is necessary as the aim, the dipole
coupling, is also non-covariant. Terms like $\sqrt{-g}$ or
$\sqrt{g_\Sigma}$ ($g_{\Sigma}$ is the determinant of the metric
on the hypersurface $x^0=$ const.) are
not included in the integrand of Eq.~(\ref{switchgl}) because
this would destroy the derivation if the definition of $\vec{P}$
is not appropriately modified. Such alternative derivations
should lead to the same result if measurable quantities are
considered.

It is not difficult to show that
the (flat space) polarization field $\vec{P}$
is related in the sense of
distributions to the (flat space) magnetization
\begin{equation} \vec{M}(y) = \sum_{\alpha} q_{(\alpha)}\,
     (\vec{x}_{(\alpha)}
     -\vec{R}_{cm})\times (\dot{\vec{x}}_{(\alpha)}
     -\dot{\vec{R}}_{cm})\, \int_0^1 \lambda \delta (\vec{y}
     -\vec{R}_{cm}
     -\lambda (\vec{x}_{(\alpha)} -\vec{R}_{cm}))\, d \lambda
     \;\end{equation}
and the R\"ontgen current $\vec{J}_{\mbox{{\scriptsize
R\"o}}} =$ rot$ (\vec{P} \times \dot{\vec{R}}_{cm})$ via
\begin{equation} \dot{\vec{P}} + \mbox{rot} \vec{M} + \vec{J}_{
     \mbox{{\scriptsize R\"o}}} =
     \sum_{\alpha} q_{(\alpha)}\, \dot{\vec{x}}_{(\alpha)}\,
     \delta (\vec{y} - \vec{x}_{(\alpha)}) - Q
     \dot{\vec{R}}_{cm}
     \delta (\vec{y} -\vec{R}_{cm}) \label{ersetz} \; .
     \end{equation}
In the remainder we assume the system to be neutral, $ Q \equiv
\sum_{\alpha} q_{(\alpha)} = 0$. In addition, we decompose
vectors $\vec{V} = \vec{V}^\| + \vec{V}^\bot$ related to the
electromagnetic field into
its longitudinal (rot $\vec{V}^\| =0$) and transverse
( div $\vec{V}^\bot =0$) part, for example
\begin{equation} \vec{A} = \vec{A}^\| + \vec{A}^\bot \; ,\;
     \mbox{rot}
     \vec{A}^\| =\mbox{div}\vec{A}^\bot = 0 \; . \end{equation}
Note that the spatial integral over the scalar product of any
transverse vector with any longitudinal vector vanishes and
that the longitudinal part of the polarization for $Q=0$
is given by
\begin{equation} \vec{P}^\| = \nabla V_0 \end{equation}
where
\begin{equation} V_0(\vec{x}) = - \frac{1}{4 \pi} \int d^3y
     \frac{\mbox{div} \vec{P}(\vec{y})}{|\vec{x}-\vec{y}|}
     = \frac{1}{4\pi} \sum_{\alpha} \frac{q_{\alpha}}{|\vec{x}
     - \vec{x}_{(\alpha)}|} \; . \end{equation}
With Eq. (\ref{ersetz}) we arrive at
\begin{eqnarray} L^\prime &=& \sum_{\alpha} \Bigg \{
     {m_{(\alpha)} \over 2}
     h_{00} + m_{(\alpha)}\, \vec{h}_0 \cdot  \dot{
     \vec{x}}_{(\alpha)} + {m_{(\alpha)} \over 2} \left
     [(1+{1\over
     2} h_{00})\delta_{kl} + h_{kl}\right ] \dot{x}^k_{(\alpha)}
     \dot{x}^l_{(\alpha)} + q_{(\alpha)}\, A_0 (x_{(\alpha)})
     \Bigg \} \nonumber \\ & &
     + {1\over 2} \int d^3 y \Bigg \{ \left [ (1+{1\over 2}
     h_{\lambda \lambda}) \delta_{kl} -\! h_{kl} \right ]
     (A_{k,0}\! - \! A_{0,k})(A_{l,0}\! -\! A_{0,l}) \! - \!
     \left [ (1\! -\! {1\over
     2} h_{\lambda \lambda}) \delta_{kl} +h_{kl} \right ]
     B_k B_l \nonumber \\ & &
     - 2(\dot{\vec{A}} - \nabla A_0) \cdot (\vec{B}
     \times \vec{h}_0) \Bigg \} + \int d^3 y \Big \{
     \vec{A}^\bot \cdot [ \mbox{rot}
     \vec{M} + \vec{J}_{\mbox{{\scriptsize R\"o}}}]
     -\dot{\vec{A}}^\bot \vec{P}^\bot - \dot{\vec{A}}^\|
     \vec{P}^\| \Big \}  \label{lag2} \; . \end{eqnarray}
For notational convenience we have introduced the vector $\vec{
h}_0 \equiv \{ h_{0i} \}$. The new Lagrangian $L^\prime$ has
the feature that it depends only on $\dot{A}_i^\| $, not on
$A_i^\|$ which is therefore a cyclic variable. The
corresponding Routhian (see, e.g., Ref. \cite{goldstein}) is
given by
\begin{equation} R = L^\prime - \int d^3 y \dot{A}^\|_i
     \frac{\partial
     {\cal L}^\prime}{\partial \dot{A}_i^\|} \; . \end{equation}
The derivative in the r.h.s. is given by
\begin{equation} \frac{\partial {\cal L}^\prime}{\partial
    \dot{A}_i^\|} = - \Delta_i^\| \end{equation}
which is a constant of motion. The vector $\vec{\Delta}$ is defined
by
\begin{equation} \Delta_i := - \left [ \left (1 + {1\over 2}
     h_{\lambda \lambda}\right ) \delta_{ij} -h_{ij} \right ]
     (A_{j,0} - A_{0,j}) + (\vec{B}\times \vec{h}_0)_i + P_i
     \label{dpara} \; . \end{equation}
and agrees in absence of a gravitational field with the
electric displacement. Using Eq. (\ref{dpara}) to eliminate
the cyclic variables from $R$ one finds an expression with no
time derivative of $A_0$ and which depends linearly on it.
Hence, $A_0$ plays the role of a Lagrangian multiplier.
Solving the corresponding constraint leads to
$ \vec{P}^\| - \vec{\Delta}^\|  = \nabla V_0 $ which implies
$ \vec{\Delta}^\| = 0$. We thus find for the Routhian
\begin{eqnarray} R &=& \sum_{\alpha} \Bigg \{ {m_{(\alpha)}
     \over 2} h_{00} + m_{(\alpha)}\, \vec{h}_0 \cdot  \dot{
     \vec{x}}_{(\alpha)} + {m_{(\alpha)} \over 2} \left
     [(1+{1\over
     2} h_{00})\delta_{kl} + h_{kl}\right ] \dot{x}^k_{(\alpha)}
     \dot{x}^l_{(\alpha)} \Bigg \} -V_{coul}
     \label{routhian} \\ & &
     + {1\over 2} \int d^3 y \Bigg \{ (\dot{\vec{A}}^\bot)^2 -
     2 \vec{P}^\bot \cdot \dot{\vec{A}}^\bot + 2 \vec{B}\cdot
     (\vec{M} + \vec{P}\times \dot{\vec{R}}) - \!
     \left [ (1\! -\! {1\over 2} h_{\lambda \lambda})
     \delta_{kl} +h_{kl} \right ]B_k B_l  \nonumber \\ & &
     - 2(\vec{P}^\| +
     \dot{\vec{A}}^\bot ) \cdot (\vec{B} \times \vec{h}_0) +
     \left [ {1\over 2} h_{\lambda \lambda} \delta_{kl}-h_{kl}
     \right ] (\vec{P}^\| + \dot{\vec{A}}^\bot )_k
     (\vec{P}^\| + \dot{\vec{A}}^\bot
     )_l \Bigg \} \nonumber  \end{eqnarray}
where
\begin{equation} V_{coul} = \frac{1}{8\pi}\sum_{\alpha,\beta}
     \frac{q_{\alpha} q_{\beta}}{|\vec{x}_{(\alpha)} - \vec{x}
     _{(\beta)}| } = {1\over 2} \int d^3 y (\vec{P}^\|)^2
     \label{coulgl} \end{equation}
is the total Coulomb interaction between all charges.
One may interpret this explicit occurence of the Coulomb
interaction between the particles as an indication that the
textbook wavefunctions for atomic electrons, which are usually
derived by assuming the interaction to be of the Coulomb-type,
belong to the dipole coupling.

The Hamiltonian is defined by the usual relation
\begin{equation} H = \sum_{\alpha =1}^N \vec{p}_{(\alpha)} \cdot
     \dot{\vec{
     x}}_{(\alpha)} + \int d^3 y \vec{\Pi}^\bot \cdot \dot{
     \vec{A}}^\bot - R \end{equation}
where $\vec{p}_{(\alpha)}$ and $\vec{\Pi}^\bot$ are the
canonical momenta of the particles and the transverse
electromagnetic field, respectively. Performing the calculations
and neglecting terms of the order $h_{\mu \nu}/m_{\alpha}$
we arrive at
\begin{eqnarray} H &=& \sum_{\alpha=1}^N m_{\alpha}\left ( 1
     -{1\over 2} h_{00}(x_{(\alpha)}) \right ) +
     \sum_{\alpha=1}^N \frac{1}{2 m_{\alpha}} \left ( \vec{p
     }_{(\alpha)} - m_{\alpha} \vec{h}_0 (x_{(\alpha)}) -
     \int d^3y \vec{B}\times \vec{n}_{\alpha} \right )^2
     \nonumber \\
     & & + {1\over 2} \int d^3y \Bigg \{ \left [ (1 \! - \!
     {1\over 2}
     h_{\lambda \lambda} ) \delta_{kl} + h_{kl} \right ] \Big [
     (\Delta^\bot_k \Delta^\bot_l + B_k B_l ) -2 \Delta^\bot_k
     P_l + P_k P_l \Big ] \nonumber \\ & &
     + 2 (\vec{P}-\vec{\Delta}^\bot) \cdot (\vec{B}\times
     \vec{h}_0) \Bigg \}  \label{ergham1}\end{eqnarray}
where the vectors $\vec{n}_{\alpha}$ are defined by
\begin{equation} \vec{n}_{\alpha}(\vec{y}) :=
     \frac{m_{\alpha}}{M}
     \left [\vec{P} (\vec{y})\!  - \! \sum_{\beta}
      q_{\beta}\vec{r}
     _{(\beta)} \int_0^1 \! \! \lambda \delta (\vec{y}
     -\vec{R}_{cm}
     -\lambda \vec{r}_{(\beta)})\,  d \lambda \right ]
     + q_{\alpha} \vec{r}_{(\alpha)} \int_0^1 \! \! \lambda
     \delta
     (\vec{y} -\vec{R}_{cm} -\lambda \vec{r}_{(\alpha)})\,  d
     \lambda \; . \end{equation}
In the last equation we have introduced the nonrelativistic
relative coordinates and the center of mass position
\begin{equation} \vec{r}_{(\alpha)} := \vec{x}_{(\alpha)} -
     \vec{R}_{cm}\;
     , \quad \vec{R}_{cm} := \frac{1}{M} \sum_{\alpha=1}^N
     m_{\alpha}\vec{x}_{(\alpha)} \end{equation}
which will be of use in the derivation of the dipole
approximation in the following section.

Note that this Hamiltonian contains explicitly the
(modified) Coulomb potential. This can be seen by using
Eq.~(\ref{coulgl}) for the evaluation of the term proportional
to $P_k P_l$ in Eq.~(\ref{ergham1}); we will discuss
the modifications to the Coulomb potential below.
The term proportional to $\Delta^\bot_k P_l$ describes the
modified dipole coupling between the transverse electric
displacement and the total dipole momentum of the atom.
\section{The dipole approximation}
To gain more physical insight into the result (\ref{ergham1})
it is of advantage to perform the dipole approximation which
is valid for many quantum optical applications. In our context
it amounts in the restriction to terms which are at most
linear in the relative coordinates $\vec{r}_{(\alpha)}$.
In this limit the polarization and the magnetization become
\begin{equation} \vec{P}(y) = \vec{d} \delta (\vec{y}
     -\vec{R}_{cm}) \; , \;
     \vec{M}(y) = {1\over 2} \sum_{\alpha} q_{(\alpha)}\,
     \vec{r}_{(\alpha)}\times \dot{\vec{r}}_{(\alpha)} \,
     \delta (\vec{y}-\vec{R}_{cm})  \end{equation}
where
\begin{equation} \vec{d} := \sum_{\alpha=1}^N q_{\alpha}
     \vec{r}_{(\alpha)} \end{equation}
is the dipole operator of the atom.
It is convenient to exploit the fact that the Hamiltonian
transforms as a scalar under a change of the dynamical
variables. We take as new variables of the particles the
center of mass position $\vec{R}_{cm}$ and the relative
coordinates $\vec{r}_{(\alpha)}$ of the first $N-1$ particles.
Denoting the corresponding canonical momenta as $\vec{P}_{cm}$
and $\hat{\vec{p}}_{(\alpha)}$ it is not difficult to show that
these are related to the old momenta by
\begin{eqnarray} \vec{p}_{(\alpha)} &=& \frac{m_{\alpha}}{M}
     \left (\vec{P}_{cm} - \sum_{\beta=1}^{N-1} \hat{\vec{p}}_{
     (\beta)} \right ) + \hat{\vec{p}}_{(\alpha)} \; , \quad
     \alpha = 1, \ldots , N-1 \nonumber \\
     \vec{p}_N &=& \frac{m_N}{M} \left
     (\vec{P}_{cm} - \sum_{\beta=1}^{N-1} \hat{\vec{p}}_{
     (\beta)} \right ) \; . \end{eqnarray}
Inserting this into Eq.~(\ref{ergham1}) and performing the
dipole approximation we can write the Hamiltonian as the sum
\begin{equation} H = H_{\mbox{\scriptsize at}} +
     H_{\mbox{\scriptsize
     cm}} + H_{\mbox{\scriptsize R\"o}} + H_{
     \mbox{\scriptsize rad}} + H_{\mbox{\scriptsize int}}
     \label{dipham} \end{equation}
of five parts. The first is the internal Hamiltonian
\begin{eqnarray} H_{\mbox{\scriptsize at}} &=&
     \sum_{\alpha=1}^{N-1}
     \frac{1}{2m_{(\alpha)}} \Big ( \hat{\vec{p}}_{(\alpha)}
     - \frac{m_{(\alpha)}}{M} \sum_{\beta=1}
     ^{N-1} \hat{\vec{p}}_{(\beta)} \Big )^2 + V_{coul}
     - h_{0k,l}(\vec{R}_{cm})\, \sum_{\alpha=1}^{N-1} r^l_{(
     \alpha)} \hat{p}_{(\alpha)k}
     \nonumber \\ & &
     + {1\over 2} \int d^3y \Bigg \{ (\vec{P}^\bot)^2 -
     \left [{1\over2}h_{\lambda \lambda} \delta_{kl}- h_{kl}
     \right ]\Big ( P_k^\bot P_l^\bot + P^\|_k P^\|_l
     + 2 P^\bot_k P^\|_l \Big ) \Bigg \}
     \label{haat} \end{eqnarray}
which does not depend on the transversal electromagnetic field
and $\vec{P}_{cm}$. The first sum is the kinetic energy term in
which the subtraction of the sum over $\beta$ describes
the generalization of the reduced mass for more than two
particles. The last term in the second line contains
modifications to the Coulomb potential and to the dipole energy.
It should be noted that the Coulomb
potential contains the self-energy of the particles and is
therefore divergent. This is also the case for the term
proportional to $(\vec{P}^\bot)^2$
which describes the dipole self-energy.
The last sum in the first line describes the coupling of
the internal angular momentum to a rotation. This can be seen
by switching to the Fermi coordinates of an rotating observer
in the weakly curved space \cite{kpm94}. In this coordinate
system the components $h_{0i}(\vec{x})$ are essentially given
by $\varepsilon_{ijk} \omega^j x^k$ where $\omega^l$ is the
angular velocity of the observer. Inserting this into
Eq.~(\ref{haat}) shows that the corresponding term is of the
form $ - \vec{\omega} \cdot \sum_{\alpha}
\vec{L}_{(\alpha)}$ if $\vec{L}_{(\alpha)}$ is the orbital
angular momentum of particle $\alpha$.

The center of mass contribution
\begin{equation} H_{\mbox{\scriptsize cm}} = \frac{1}{2M}
     \vec{P}_{cm}^2 -
     \frac{M}{2} h_{00}(\vec{R}_{cm}) - \vec{P}_{cm}\cdot
     \vec{h}_0 (\vec{R}_{cm}) \end{equation}
has the same structure as the Hamiltonian for a free particle
in a weakly curved space, comp. Refs.~\cite{dewitt66,papini67}.
The internal and external degrees of freedom of the atom are
coupled via the R\"ontgen term
\begin{equation} H_{\mbox{\scriptsize R\"o}} = - \frac{1}{M}
      \vec{P}_{cm}
      \cdot (\vec{B} \times \vec{d}) \; . \end{equation}
This is the same expression as in Minkowski space
because we have neglected all terms of the order
$h_{\mu \nu} / m_{(\alpha)}$. The radiative part
of $H$ is found to be
\begin{equation} H_{\mbox{\scriptsize rad}} = \frac{1}{2} \int
     d^3 y \Bigg \{ \left
     [ \left ( 1-{1\over 2} h_{\lambda \lambda} \right ) \delta
     _{kl}  + h_{kl} \right ] \left (  \Delta_k
     ^\bot \Delta_l^\bot + B_k B_l \right )
     - 2 \vec{h}_0 \cdot (\vec{\Delta}^\bot \times \vec{B})
     \Bigg \} \end{equation}
and contains a coupling between the Poynting vector and the
rotation. Here we have used  the relation $\vec{\Pi}^\bot =
-\vec{\Delta}^\bot$.

The interaction between matter and radiation is described by
\begin{equation} H_{\mbox{\scriptsize int}} = - \left [\left (
     1-{1\over 2}
     h_{\lambda \lambda} \right ) \delta_{kl}+ h_{kl} \right ]
     \Delta^\bot_k d_l - \vec{B} \cdot \Bigg \{
     \sum_{\alpha = 1}^{N-1} { q_{(\alpha)}\over 2 m_{(\alpha)}
     } \left ( \vec{r}_{(
     \alpha)} \times \left ( \hat{\vec{p}}_{(\alpha)} -
     \frac{m_{(\alpha)}}{M} \sum_{\beta=1}^{N-1} \hat{\vec{p}}
     _{(\beta)} \right ) \right )
     \Bigg \} \; . \label{hdip}\end{equation}
The first term describes the dipole coupling between the atom
and the transverse electromagnetic field. The second term is
the well known coupling between the magnetic field and the
angular momentum of the particles. As in Eq.~(\ref{haat})
the mass reduction has to be taken into account. Again this
term is the same as in flat space since we have neglected
terms of the order of $O(h_{\mu \nu} / m_{(\alpha)})$.
\section{The dipole coupling between measurable quantities}
To make contact with the measurable quantities of the theory
it is necessary in general relativity to consider the
components of each tensor with respect to a tetrad field
$e_{\underline{\alpha}}^\mu$ which fulfills
$e_{\underline{\alpha}}^\mu e_{\underline{\beta} \mu} =
\eta _{\underline{\alpha} \underline{\beta}}$ at each point in
space. Here $\eta_{\underline{\alpha} \underline{\beta}}$ is
the Minkowski metric. The measured components of, e.g., the
electric field are then given by
\begin{equation} E^{\underline{i}}=
     F^{\underline{0i}} = e^{\underline{0}\mu} e^{\underline{
     i}\nu} F_{\mu \nu}  \end{equation}
(Tetrad indices are raised and lowered with the Minkowski
metric).
We now focus on the dipole term in Eq.~(\ref{hdip}). The
vector $e_{\underline{0}}$ of the tetrad is assumed to be
orthogonal to the hypersurfaces $x^0 = $const.~and is therefore
given by $e_{\underline{0}}^0 = 1 + h_{00}/2$, the rest of its
components vanish. The orthogonality of the tetrad then implies
$e_{\underline{i}0} =0 $ for all three space-like
vectors $e_{\underline{i}}$.

Consider now the dipole term in Eq.~(\ref{hdip}) and insert
Eq.~(\ref{dpara}). The dipole coupling then has the form
\begin{equation} - \left [- F_{0l} + (\vec{B}\times \vec{h}_0)_l
     + \left [\left ( 1-{1\over 2} h_{\lambda \lambda} \right )
     \delta_{kl}+ h_{kl} \right ] P_k \right ]^\bot d_l \; .
     \end{equation}
Since the longitudinal part of the vector $\Delta$ is zero we can
omit
in this expression the index $\bot$. First we examine the term
proportional to $F_{0l}$.
Recalling that the dipole moment is the sum of relative
coordinates\footnote{To achieve
true invariant quantities it would be
necessary to work with the derivative
$\partial_{\mu}(s_{(\alpha)}^2)/2$
of the geodesic distance $s_{(\alpha)}$
between the center of mass
of the atom and the particle $\alpha$ instead of the relative
coordinate $\vec{r}_{(\alpha)}$ of the particle. But since we
are working in the dipole approximation (up to linear terms
in $\vec{r}_{(\alpha)}$) we have $r_{(\alpha)}^\mu \approx
g^{\mu \nu} (R_{cm}) \, \partial_{\mu}(s_{(\alpha)}^2)/2 $
so that the difference is unimportant.}
$r^l_{(\alpha)}$ it is not difficult to find
\begin{eqnarray} F_{0l} \sum_{\alpha} q_{(\alpha)}
     r_{(\alpha)}^l &=&
     (1- h_{00}/2)F_{\underline{0l}}\sum_{\alpha}
     q_{(\alpha)}  r_{(\alpha)}^{\underline{l}} \nonumber \\
     &=& -(1-h_{00}/2) E_{\underline{l}} \sum_{\alpha}
     q_{(\alpha)} r_{(\alpha)}^{\underline{l}}   \end{eqnarray}
If we interpret
\begin{equation} P^G_l := (1+ h_{00}/2) \left [\left
     ( 1-{1\over 2} h_{\lambda \lambda} \right )
     \delta_{kl}+ h_{kl} \right ] P_k + (\vec{B}\times
     \vec{h}_0)_l \end{equation}
as the actual polarisation field modified by the gravitational field
we find for the dipole term the expression
\begin{equation} H_{dip} = - (1- h_{00}/2)
     D_{\underline{l}}^\bot d^{\underline{l}}
     \; . \end{equation}
Compared to flat space the dipole coupling seems to be modified
by the factor $(1- h_{00}/2)$ which is identical to $\sqrt{-g_{00}}$.
But even this term disappears if we take into account that the
Hamiltonian (\ref{dipham}) describes the time evolution of the
system with respect to the coordinate time $x^0$. If we switch to the
proper time $\tau$ of a family of observers with
four-velocity $u^\mu = e_{\underline{0}}^\mu$ the Hamiltonian
density ${\cal H}$,
defined by $H = \int {\cal H} d^3y$, has to be transformed
according to
\begin{equation} {\cal H}^\prime = {\cal H} { dx^0 \over d \tau}
     = {\cal H}
     e^0_{\underline{0}} = {\cal H} {1 \over \sqrt{-g_{00}}} \;
     . \label{zeitwe} \end{equation}
This implies that the dipole coupling between the atoms and the
electric field  takes the form
\begin{equation} H^\prime_{dip} = - D^\bot_{\underline{l}}
     d^{\underline{l}} \end{equation}
even in a weakly curved space when it is expressed with respect
to the measured quantities and the proper time of an observer.
This result is in agreement with a recent work of
L\"ammerzahl \cite{laemm94} in which he
derives the dipole coupling
for a quantum mechanical particle moving in a PPN space-time.

We now discuss briefly the modifications to the Coulomb
potential as this is of interest for the decision to which
form of coupling the textbook wavefunctions belong (see the
remark below Eq.~(\ref{coulgl})). It is of advantage to start
with the expression
\begin{equation} V := {1\over 2} \int d^3y \left [ (1 \! - \!
     {1\over 2}
     h_{\lambda \lambda} ) \delta_{kl} + h_{kl} \right ]
     P_k P_l  \end{equation}
in Eq.~(\ref{ergham1}) which contains the Coulomb as well
as the dipole energy of the atom. In a first step we replace
the flat space polarization $P_l$ by its gravitational
counterpart $P^G_l$ and arive at
\begin{equation} V = {1\over 2} \int d^3y\Bigg \{ (1-h_{00})
     \left [ (1 \! +
     \! {1\over 2} h_{\lambda \lambda} ) \delta_{kl} - h_{kl}
     \right ] P^G_k P^G_l - 2 \vec{P}^G \cdot (\vec{B}\times
     \vec{h}_0) \Bigg \} \; .\end{equation}
The last term can be interpreted as a gravitational induced
interaction between the atom and the magnetic field and
is therefore not a part of the Coulomb or dipole energy; we
will omit this part. Performing the same steps as for the
dipole coupling we find that the first term has, after
the insertion of the tetrad vectors, the form
\begin{equation} {1\over 2} \int d^3y (1+ h^\lambda_{\lambda}/2)
     P^G_{\underline{i}}P^{G \underline{i}} \; . \end{equation}
It only left to replace the coordinate time by the
proper time of the observers according to Eq.~(\ref{zeitwe}).
This leads to the result that the sum of Coulomb and dipole
energy is given by
\begin{equation} {1\over 2} \int dV P^G_{\underline{i}}P^{G
     \underline{i}} \end{equation}
where $dV = (1+ h_{ii}/2) d^3y = \sqrt{g_{\Sigma}} d^3y$ is
the three-volume element of the hypersurface $x^0 =$ constant.
It has therefore the same form as in flat space if it is
written with respect to the measurable quantities and with
respect to the proper time.

We thus find the result that in connection with the dipole
coupling the modifications of the Coulomb potential are
related to the modifications of the (longitudinal part of)
the polarization field. If the calculation of the
modifications in the minimal coupling scheme are different
from the present result this would, in principle, open a new
way to test experimentally to which coupling the
textbook wavefunctions belong. It should be noted, however,
that the magnitude of the modifications in atomic systems is
far too small to be measurable.

This section will be closed with some general remarks concerning
the use of the family of observers to define the tetrads
and therefore the measurable quantities. In principle it
would be more convenient to work in the atom's frame of
reference instead of including the family of
obervers in the description.
But in general relativistic situations a frame of reference
cannot be unambiguously defined. Usually one uses
Fermi coordinates \cite{manasse63} as a local approximation
for the reference frame, but this construction has some
shortcomings. A modification of it may circumvent these
problems \cite{kpm94b}, but the true form of a reference frame,
if it exists, has to be determined by experiments.

A second remark concerns the fact that the Hamiltonian is
changed if we switch to the proper time of the family of
observers. This is due to the fact that the Hamilton operator
is the time evolution operator of the system. If we change the
time coordinate then the evolution operator is also changed.
The use of a proper time is of advantage compared to the
coordinate time since it is directly measurable by the clocks
of the observers.
\section{The spin interaction}
In the derivation of the Hamiltonian
the inclusion of the spin was not possible
because only classical particles were considered.
But the heart of the Power-Zienau transformation,
the subtraction of a total time derivative from the Lagrangian,
can be made without any reference to the spin. Hence, it should
be possible to include the spin by  deriving the
minimal coupling Hamiltonian and by taking over the
spin terms into the Hamiltonian with dipole coupling.
In order to get the correct spin terms we will follow closely
the approach of Ref.~\cite{maau93b}. For brevity we will only
scetch the main steps and refer to this paper for further
details.

The Dirac equation in a weak gravitational field can be written
as
\begin{equation} i \partial_0 \psi = H \psi \end{equation}
with
\begin{eqnarray} H &=& -q A_0 +\left [ \left (1- {1\over
     2}h_{00}\right ) \alpha_{\underline{i}}
     - h_{0i} - {i\over 2} h_{0j} \varepsilon_{jik} \Sigma_{
     \underline{k}} - {1\over 2} h_{ij} \alpha_{\underline{j}}
     \right ] \, (-i \partial_i -q A_i ) \nonumber \\
     & & + {i \over 4} (h_{0i,i} -
     h_{ii,0}) + {i\over 4} \alpha_{\underline{i}} (h^\rho_{
     \; i,\rho} - h^\rho_{\; \rho ,i}) -i m \left [ \left (1-
     {1\over 2} h_{00}\right )\gamma_{\underline{0}} -{1\over
     2}h_{0i}\gamma_{\underline{i}} \right ] \end{eqnarray}
In order to give the scalar product between spinors the usual
form in flat space we redefine the field by
\begin{equation} \psi = O \psi^\prime \mbox{ with } O = 1
     -{1\over 4} h_{ii} - {1\over 4} h_{0i}
     \alpha_{\underline{i}} \end{equation}
The corrections to the Pauli equation can be found by
performing a Foldy-Wouthuysen transformation (see, e.g.,
Ref.~\cite{ItZu}) with the unitary operator $\exp (iS)$ where
the Hermitean operator $S$ is given by
\begin{equation} S= {1\over 2m} \gamma_{\underline{i}} (-i
     \partial_i -qA_i)
     -{1\over 4m} h_{ik} \gamma_{\underline{k}} (-i \partial_i
     -qA_i ) +{i\over 8m} \gamma_{\underline{i}} h_{ki,k}
     \end{equation}
The Hamiltonian for the Schr\"odinger field is then found to be
\begin{equation} H={1\over 2m} (-i \partial_i -qA_i -m h_{0i})^2
     -q A_0
     +m\left (1-{1\over 2} h_{00}\right ) -{1\over 4}h_{0l,i}
     \varepsilon_{ilk}\sigma_k -{q\over 2m} B_k \sigma_k
     \end{equation}
This is the same result as in Refs.~\cite{dewitt66,papini67}
except for the inclusion of the spin-gravity interaction
\begin{equation} -{1\over 4}h_{0l,i} \varepsilon_{ilk}\sigma_{k}
     \; . \end{equation}
Here $\sigma_{k}$ are the
Pauli matrices  of the particle under consideration, say
particle $\alpha$. In the discussion of the result
(\ref{dipham}) we have seen that the term $h_{0l,i}$ is
essentially given by $ \varepsilon_{lmi} \omega^m$ with
the angular velocity $\omega^m$ of the observer. It is therefore
obvious that this term describes the $- \vec{\omega} \cdot
\vec{S}$ coupling of the spin part $\vec{S}$ of the total angular
momentum to the rotation $\vec{\omega}$.

The derivation along the lines of Ref. \cite{maau93b} shows
also that the Hamiltonian is only related to a relativistic
Hermitean energy operator if the gravitational field is
time independent. Since the spinless part of it is exactly
the same as the canonical Hamiltonian, and since the
canonical procedure is not manifest covariant, we conclude that
the same condition must hold for  the Hamiltonian (\ref{dipham}).
\section{Discussion}
In this paper we have shown that the interaction between atomic
particles and the electromagnetic field can be described with
a dipole coupling term also if space-time is weakly curved.
The form of the coupling remains the same as in Minkowski space
if it is written with respect to the proper time of an observer
and the measurable quantities of the theory. The same is true
for the Coulomb and the dipole energy of the atom.

The central
assumptions in this derivation are the smallness of the
velocity of each particle, the weakness of the gravitational
field, and the validity of the dipole approximation.
A further assumption was tacitly made by using the
non-relativistic form of the center-of-mass coordinates.
Although this should in general be a good
approximation, Fischbach {\em et.~al.} \cite{fischbach81}
have shown that the use of different relative coordinates
(center-of-energy, e.g.) can result in different
perturbational contributions of the weak gravitational
field which do not vanish even when the mass of the nucleus
is very large. We assume that these contributions are small
enough to be neglected.

It is interesting to make a comparison of the present results
with the well known formal equivalence between the Maxwell-field
in a curved space and a dielectric medium \cite{volkov71}.
In this approach one defines a formal dielectric displacement
vector to describe the influence of gravity on the Maxwell
field. In absence of particles, i.e. for vanishing polarisation
$\vec{P}$, the formal electric displacement agrees with
the vector $\vec{\Delta}$ defined above (in Ref.~\cite{volkov71}
the presence of charged matter was not considered).
Also the coupling of the Poynting vector to the rotation
occurs in the energy density of the formal Maxwell field.
\\[5mm]
{\bf Acknowledgement}\\
I would like to thank J.~Audretsch for helpful remarks,
C.~L\"ammerzahl for very fruitful discussions on
measurable quantities, and the Studienstiftung des Deutschen
Volkes for financial support.
\nopagebreak

\end{document}